\documentclass[aip, jcp, preprint, groupedaddress]{revtex4-1}

\usepackage{graphicx}
\usepackage{amsmath}
\usepackage{color}
\usepackage{threeparttable}
\usepackage{appendix}
\usepackage[normalem]{ulem}
\usepackage{array,multirow}
\usepackage{epstopdf}

\begin{document}

\def\deps{\Delta\epsilon_g}
\def\PBE{^{\rm PBE}}
\def\n{n}
\def\app{^{\rm app}}
\def\bea{\begin{eqnarray}}
\def\eea{\end{eqnarray}}
\def\ben{\begin{equation}}
\def\een{\end{equation}}
\def\sss{\scriptscriptstyle\rm}
\def\HF{^{\rm HF}}
\def\x{_{\sss X}}
\def\c{_{\sss C}}
\def\s{_{\sss S}}
\def\xc{_{\sss XC}}
\def\dc{_{\sss DC}}
\def\ext{_{\rm ext}}
\def\ee{_{\rm ee}}
\def\sint{ {\int d^3 r \,}}
\def\br{{\mathbf r}}
\def\K#1{{\bf #1}}

\title{Ions in solution: Density Corrected Density Functional Theory (DC-DFT)}

\author{Min-Cheol Kim}
\author{Eunji Sim$^*$}
\affiliation{Department of Chemistry and Institute of Nano-Bio Molecular Assemblies, 
Yonsei University, 50 Yonsei-ro Seodaemun-gu, Seoul 120-749 Korea}

\author{Kieron Burke}
\affiliation{Department of Chemistry, University of California, Irvine, CA, 92697, USA}

\date{\today}


\begin{abstract}

Standard density functional approximations often give questionable results for odd-electron radical complexes, with the error typically attributed to self-interaction.  In density corrected density functional theory (DC-DFT), certain classes of density functional theory calculations are significantly improved by using densities more accurate than the self-consistent densities. We discuss how to identify such cases, and how DC-DFT applies more generally. 
To illustrate, we calculate potential energy surfaces of HO$\cdot$Cl$^-$ and HO$\cdot$H$_2$O complexes using various common approximate functionals, with and without this density correction. Commonly used approximations yield wrongly shaped surfaces and/or incorrect minima when calculated self consistently, while yielding almost identical shapes and minima when density corrected.  This improvement is retained even in the presence of implicit solvent. 

\end{abstract}

\maketitle

\setcounter{secnumdepth}{2}
\renewcommand\thesubsection{\Alph{subsection}}

\section{Introduction}

Odd-electron radical complexes like HO$\cdot$Cl$^{-}$ and HO$\cdot$H$_2$O are of tremendous importance throughout chemistry and in related fields, such as in radiation chemistry, atmospheric chemistry, environmental chemistry, and cell biology\cite{IntroRadiatChem, GDC05, AtmosChemPhys, BGH88, POM06, POM07, RadBioMed}. In particular, the behavior of anions in droplets is of critical importance to understanding atmospheric chemistry.  Common sense suggests that anions are less perfectly screened near a water-air interface, and so have lower concentration there. Recent classical molecular dynamics (MD) simulations have shown just the opposite\cite{DKT08}, creating  considerable controversy on this point.  Since anions are strongly quantum mechanical, it is logical to check traditional MD simulations, using only classical force fields, against {\em ab initio} molecular dynamics (AIMD) calculations, which use density functional theory (DFT) to generate the potential energy surfaces (PES).

Unfortunately, standard DFT approximations have issues for such systems\cite{GKC04, GKC04b, PZ02, ACCS12}. Several studies show DFT approximations predict two minima in the ground-state PES. One of the minima is a hydrogen-bonding structure, while the other is a two-center three-electron interacting hemi-bonding structure\cite{C11}. Many DFT studies predict the hemi-bonding structure as the global minimum of [HO$\cdot$Cl(H$_2$O)$_{n}]^-$ complex\cite{B93, VLB05, DKT08, VVS05, Y11}. This is attributed to the infamous self-interaction error\cite{DKT08, VVS05, Y11}, because AIMD studies with self-interaction corrected (SIC) methods agree with PES scans with high-level (beyond DFT) quantum-chemical methods, showing no hemi-bonding configuration in the ground state\cite{DKT08, VVS05, Y11}. Hemi-bonding configurations have been observed in experiments\cite{CS69, SSE97}, but high-level quantum-chemical studies show that these are excited-state rather than ground-state configuration\cite{VDT09}.

 Thus, high-level quantum chemical calculations reveal that the ground-state PES has only one minimum, which is the hydrogen-bonding structure. The hemi-bonding structure is relatively overstabilized in DFT because three electrons are incorrectly delocalized over two atoms rather than having localized electron configurations.  Essentially, the AIMD studies raise more questions than they answer.

However, recent work\cite{KSB13} produces a much simpler and more general picture of such errors, and suggests alternative solutions.  In any approximate DFT calculation, the density is found by self consistently solving the Kohn-Sham (KS) equations, using an approximate KS potential derived from the energy approximation. This procedure finds the density that minimizes the energy approximation for the given system. The final output energy is of the approximate energy functional evaluated at the approximate density.
For most KS-DFT calculations using standard approximations, the KS potential appears to be of very poor quality\cite{UG94}. In particular, the eigenvalues are usually too shallow by several eV\cite{KK08}.  DFT approximations are designed to yield good energies, but this does not automatically imply good potentials. Functional derivatives are determined by how well the approximation performs for small variations {\em away} from the density of interest.  Nevertheless, usually the self-consistent density is rather good, because the approximate KS potential is rather close to a simple shift of the exact KS potential. (Highlighting this point, a recent approximation yields highly accurate energies with truly terrible potentials\cite{SMHB13}.)

 This then raises a simple issue.  One can think of two distinct sources of error in such a calculation:  One of the error due to the energy approximation itself, and the other error in the corresponding approximate density.  In most DFT calculations, it is (correctly) believed that the first (functional) error dominates.  But in many interesting cases that are specific to the system and the approximation, the error in the approximate density can be unusually large, so that it contributes more to the total energy error than would usually be the case.  In such cases of density-driven errors, use of a more accurate density should reduce the final error substantially.  A small HOMO-LUMO gap ($\Delta\epsilon_{g}$) in the original approximate DFT calculation is often a strong indicator of a large density-driven error\cite{KSB13, GBMM12}. In many cases, with standard DFT approximations, a Hartree-Fock (HF) calculation provides a sufficiently accurate density to garner much of the improvement in energetics. We dub such a calculation, which is quite general, density corrected DFT (DC-DFT), meaning any DFT calculation in which the density is {\em not} found self consistently, but by some other method that substantially reduces the density-driven error.

We note that a better term for self-interaction error is delocalization error, which has been recently much studied by Cohen et al\cite{CMY07}.  Our analysis complements theirs.  In that language, we are able to determine if a given delocalization error is density-driven or not.  If the former, a better density will improve the energy; if the latter, as in the case of stretched H$_2^+$, it will not. We expect cases like DFT calculation on stretched He$_2^+$ to be normal, so the density-driven error would be small compared to the functional error\cite{GKC04b}.

\begin{figure*}[htb]
\begin{center}
\includegraphics[width=6.2in]{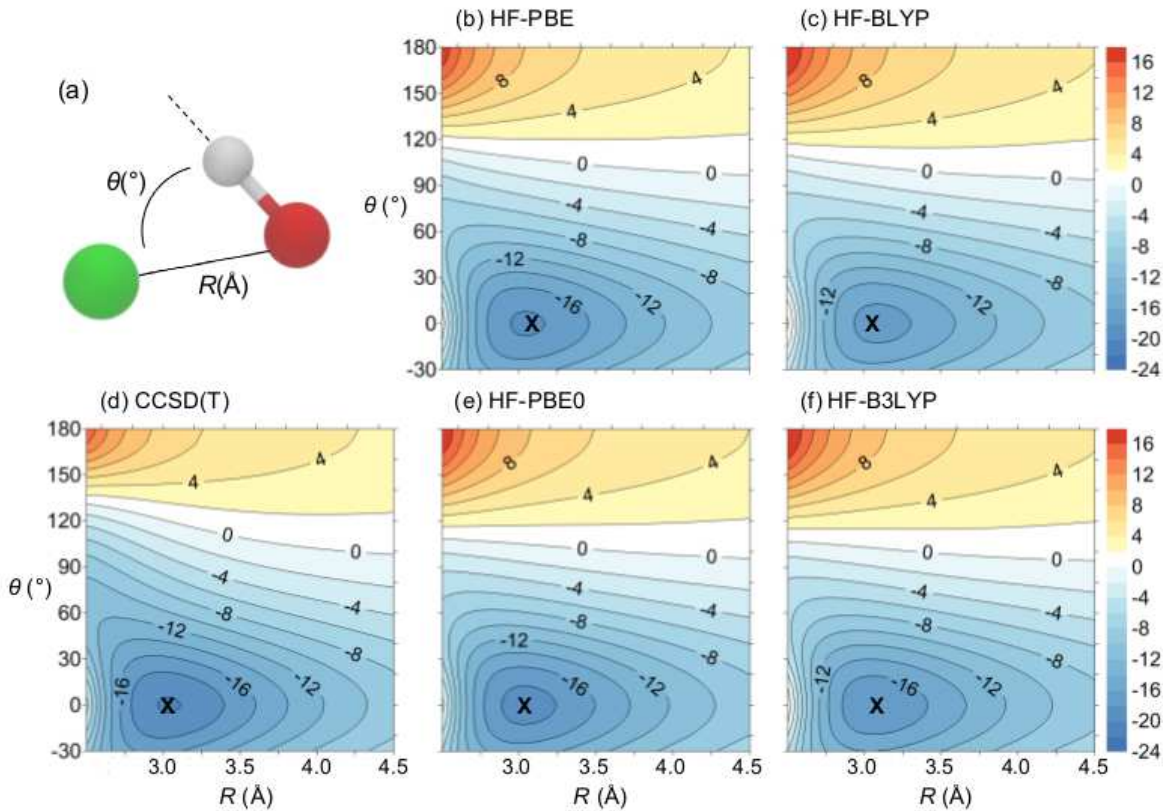}
\caption{PES of HO$\cdot$Cl$^-$ with various DFT methods on HF densities (HF-DFT), using the aug-cc-pVTZ basis set. X marks the global minimum of each PES.}
\label{HOClPESfunc2}
\end{center}
\end{figure*}

In the present work, we apply this scheme to a variety of commonly-used standard approximations in DFT, applied to the HO$\cdot$Cl$^-$ complex. We use the HF density as our more accurate density.  
We use the notation A-B, where A indicates the method for finding the density, and B the method for finding the energy. Thus HF-DFT implies using a HF density, but evaluating its energy using DFT approximation.  In Fig. \ref{HOClPESfunc2} we show the near-perfect agreement of many commonly-used functionals for the potential energy surface of HO$\cdot$Cl$^-$, once they are evaluated on HF densities. We find that not only does HF-DFT always produce a PES with the correct global minimum, but that the difference in PES's between different approximations becomes negligible and is, up to a constant, essentially identical to the CCSD(T) PES.  Thus, with this elementary correction, the PES of such systems changes from being a disappointing failure to being a resounding success for approximate DFT.

There is always a price to be paid for progress, and in this case, there are two (not so hidden) challenges to taking advantage of this improved performance. The first is that, if the density is no longer the minimizer for the given approximate energy functional, many basic theorems, such as the Hellmann-Feynman theorem, no longer apply, and many of these are used in standard DFT codes.   But the way to calculate forces has already been detailed in a pioneering work\cite{VPB12} which pointed out that HF-DFT generically improves most reaction barriers over self-consistent DFT, and implemented it in the ACES code.  The second is that AIMD simulations are often performed with plane waves enforcing periodic boundary conditions, and solving the HF equations can be computationally expensive in such implementations\cite{MPB93}.  However, some simpler scheme available in those codes, such as self-interaction corrected local density approximation (LDA-SIC)\cite{PZ81}, might yield densities sufficiently accurate for the purpose while using e.g., generalized gradient approximations (GGA) for the energy evaluation.  Alternatively with the recent success of hybrid functionals such as HSE06\cite{KVIS06} in predicting fundamental gaps, it might be straightforward to run a HF calculation under these conditions.

But all these are beyond the scope of the present paper.  The aim here is to show just how much of an improvement is possible with DC-DFT. We begin with some theory, which also includes the historical background. Since this idea is so elementary, it has been touched upon several times in the distant and recent past.  We also explain how this fits in with all the present attempts to go beyond standard approximations, including using exact exchange in DFT\cite{DG01, DG02}, {\em ab initio} DFT\cite{VB12c}, and several many-body approaches using KS orbitals\cite{F08, SHS08}. Next we show a variety of results for the complex at hand, using several functionals, basis-sets, and implicit solvent models.   We also demonstrate similar results for the HO$\cdot$H$_2$O radical, showing that the effect is not specific to anionic species, but is much more general.

\section{Theory}

\subsection{Definition of error decomposition}

In any KS-DFT calculation, only a small fraction of the total energy need be approximated as a functional of the density, namely the exchange-correlation (XC) energy.  In reality, we always use spin densities, but for the present purposes, we suppress the spin index.  The self-consistent solution of the KS equations has been designed to deliver the density that minimizes the total energy in the KS scheme\cite{WSBW13}:

\ben
\tilde E[\n] = T\s[\n] + U[\n] + V[\n] + \tilde E\xc[\n]
\een

\noindent where the $\sim$ indicates an approximation.  The functionals $E[\n]$, $T_{s}[\n]$, $U[\n]$, $V[\n]$, and $E\xc[\n]$ are the total energy, kinetic energy, external potential energy, Hartree-energy and XC energy functionals, respectively\cite{BW13}.

The energy error in a DFT calculation is defined as

\ben
\Delta E = \tilde E [\tilde\n] - E[\n]
\een

\noindent where $\tilde\n(\br)$ is the approximate self-consistent density. We may write this error as the sum of two contributions. We call the first the {\em functional} error.  It is the energy error made by the functional evaluated on the exact density, and comes entirely from the XC approximation:

\ben
\Delta E_F = \tilde E[\n] - E [\n] = \tilde E\xc[\n]-E\xc[\n].
\een

\noindent The {\em density-driven} error is the energy difference generated by having an approximate density: 

\ben
\Delta E_D = \tilde E[\tilde\n]-\tilde E[n],
\een

\noindent so that the total energy error is the sum of these two:

\ben
\Delta E = \Delta E_F + \Delta E_D.
\een

This separation applies to {\em any} approximate DFT calculation, not just a KS calculation of electronic structure. But  here we focus exclusively on the latter, since we wish to use this as a tool to analyze chemical calculations using KS-DFT. 

Note that this elementary breakdown is hardly a breakthrough. Most developers and many users of DFT have thought along these lines or come across this in some calculation or context of theory development. What {\em is} new is how far this elementary step can be taken in analyzing all practical DFT calculations, i.e., those using approximate functionals, and how to improve many of these.

\subsection{Classification of DFT calculations}

The first point to note is that for most calculations using the KS scheme and modern approximations to XC, such as a GGA or a global hybrid, the densities are remarkably accurate. The above tool allows us to specifically quantify this accuracy, by measuring it in terms of its effect on the quantity we almost exclusively care about, namely the ground-state energy. In any calculation, if $|\Delta E_D| \ll |\Delta E_F|$, any error in the density is irrelevant for practical purposes. For example, it has long been known that the density with standard approximations is highly inaccurate at large distances from the nuclei, due to the highly inaccurate HOMO in such calculations\cite{SRZ77}.  However, in most cases, this inaccuracy produces only a very small density-driven error in the energy, so such approximations remain accurate for ground-state energies.  Moreover, DFT approximations produce far more accurate ionization energies via total energy differences than via orbital energy differences. Next, we consider some popular application, such as the calculation of a bond length with DFT.  Since the bond length is extracted as the minimum of the total energy, the contributions to the error can be split into functional-driven and density-driven, and the two compared. Thus one extracts the density-driven contribution to a given property of a given system with a given approximate functional.  If this error is small or negligible compared to the actual error, we classify such a calculation as {\em normal}. We believe the vast majority of DFT calculations fit into this category, and we gave several examples in our previous work, including cases like stretched H$_2^+$ cation\cite{KSB13}.

In fact, in many circumstances of method development, there is an underlying assumption that the calculation {\em is} normal.  With a new approximate functional, it can often be the case that the functional derivative is demanding to calculate.  Thus, often a lower-level, more standard approximation is used to calculate orbitals in the KS equations, and the new approximation is tested on those orbitals.  If the calculation is normal, this (almost) guarantees that only a small error is made by this procedure, and the change upon self-consistency will be negligible.

Our main interest, naturally, will be in those calculations where the density-driven error is a significant fraction of the total.  We denote such calculations as {\em abnormal}.  In such cases, a more accurate density will reduce the error (assuming no accidental cancellation of functional- and density-driven errors).  We can also state just how much more accurate that density need be:  Enough to make the density-driven error small relative to the functional error.  In such cases, correcting the density in a DFT calculation greatly reduces the error; hence our name for this method.

While this separation scheme can be applied to any approximate DFT calculation, we are presently focused on the infamous self-interaction error inherent in standard GGA and global hybrid calculations. In earlier work, we found that many cases of such errors (small anions, underestimated transition barriers, and incorrect dissociation limits) were in fact density-driven\cite{KSB11, KSB13}.  In such cases, the density was particularly poorly described, and a simple HF density was sufficient to drive out the density-driven error.  Note that this does not mean the HF density is especially good. In the vast majority of cases (the normal ones), the HF density is worse than the self-consistent DFT density\cite{BGM13}, and HF-DFT is worse than self-consistent DFT, as we have shown in cases like stretched H$_{2}^{+}$ in our previous work\cite{KSB13}. In these normal cases, self-consistent DFT is usually sufficient enough for getting accurate energies despite having incorrect potentials\cite{KSB13, LFB10}. But for abnormal systems, the self-consistent DFT density is especially poor in a very systematic way, a way that is largely fixed by HF.  As we showed in Ref. \onlinecite{KSB13}, an unusually small KS HOMO-LUMO gap, $\deps$, in the DFT calculation indicates a likely abnormal calculation. This means the self-consistent solution is unusually sensitive to small changes in the potential, so that an error in the XC contribution to that potential can produce an unusually large effect on the density. In the case of atomic anions, this becomes extreme:  The HOMO is positive if the basis set is used to hold it in, and zero in the basis-set converged limit\cite{LFB10}. This leads to very poor densities, missing a significant fraction of an electron, and large density-driven errors.  HF densities are a major improvement in such cases.


\subsection{Simple illustration: Two-electron systems}

\begin{table}[h]
\caption{Energy decomposition of H$^{-}$ and He.}
\begin{center}
\begin{tabular}{|c |c| c c | c c c| }
\hline
\multirow{2}{*}{atom} & \multirow{2}{*}{method} & $E$ &  $\epsilon_{1 \sss s}$ & $\Delta E$ & $\Delta E_D$ & $\Delta E_F$ \\
 && \multicolumn{2}{c}{(Hartree)} \vline & \multicolumn{3}{c}{(mHartree)} \vline \\
\cline{3-7}
\hline
\multirow{2}{*}{He} & HF & -2.86 & -0.92 & 42.07 & -0.05 & 42.12 \\
& PBE & -2.89 & -0.58 & 10.82 & -0.98 & 11.80 \\
\hline
\multirow{2}{*}{H$^-$} & HF & -0.49 & -0.04 & 41.29 & -0.71 & 42.00 \\
& PBE & -0.54 & -0.00 & -10.39 & -11.40 & 1.00 \\
\hline
\end{tabular}
\end{center}
\label{2eEnergy}
\end{table}

\begin{figure}[htb]
\begin{center}
\includegraphics[width=2.5in]{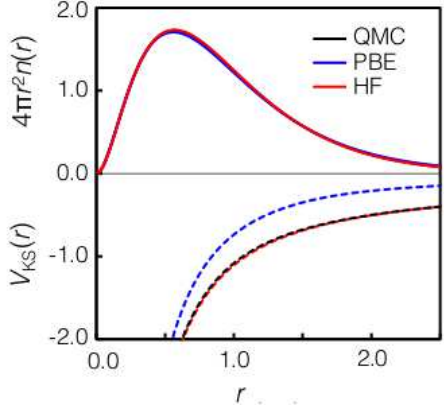}
\label{2edenpot}
\caption{Radial densities (solid line) and KS potentials (dashed line) of He
evaluated with various methods. All values are in atomic units. An atomic DFT code was used for the calculations\cite{opmks}.}
\label{Hedenpot}
\end{center}
\end{figure}


To illustrate the idea, we consider the simplest possible case, the He atom. We begin by applying our analysis to a HF calculation of the system.  For two spin-unpolarized electrons, a HF calculation is equivalent to a KS-DFT calculation with 

\ben
\tilde E\xc[\n] = - U[\n]/2.
\een

\noindent In fact, the total error in such a calculation is simply the (quantum-chemical) definition\cite{GPG96} of the negative of the correlation energy:

\ben
\Delta E = E\HF[n\HF] - E[n] = - E\c.
\een

\noindent In Table I, we list the different errors in HF and PBE calculations, for both He and H$^-$. We see that, for HF applied to He, the density-driven error is minuscule (0.05 mH), i.e., about 0.1\% of the functional error.   This says that, for this problem, the HF density is extremely accurate, and essentially all the error comes from the missing correlation energy.  Such a calculation could be considered ultra-normal.

Next, we repeat the analysis with the PBE approximation\cite{PBE96}.  Here the total error is smaller (about 11 mH), and the density-driven error is -1 mH.  Because this is still only of order 10\%, this remains a normal calculation, just like the HF calculation.   But in Fig. \ref{Hedenpot}, we show the corresponding densities and KS potentials for the PBE and HF calculation. Although the densities are identical to the eye, we see that the KS potential of the PBE calculation is far too shallow.  This is typical of all approximate DFT calculations, and leads to KS eigenvalues that are far too shallow(-0.58 eV instead of -0.903 eV).  Nevertheless, we emphasize that these are normal calculations, and the error in the potential produces very little error in the density, and so relatively small density-driven errors. We also emphasize that normality (or otherwise) is a characteristic of a particular calculation (i.e., the system and the approximation together).  

\begin{figure}[htb]
\begin{center}
\includegraphics[width=2.5in]{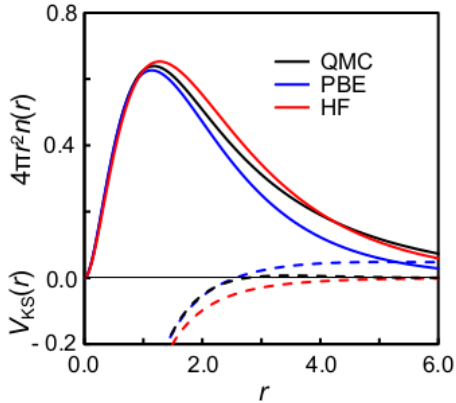}
\caption{Same as Fig. \ref{Hedenpot}, but for H$^-$.}
\label{H-denpot}
\end{center}
\end{figure}

Now we repeat this exercise for the $H^-$ ion. It is often said that H$^-$ is `correlation bound', meaning that in HF theory, H$^-$ is not stable. In a HF calculation, the ground-state energy of H is lower than that of H$^-$.  Nonetheless, we can still converge the calculation and obtain the ground-state energy.  The correlation energy remains about the same, but the density-driven error is much larger.  In Fig. \ref{H-denpot}, we can see the error in the density by eye.  Nonetheless, this calculation is also normal, with a density-driven error that is only about 3\% of the total.  It is the functional error that makes H$^-$ correlation bound.

A very different story is seen for the PBE calculation on the same system.  In fact, only by allowing a fraction of an electron (about 0.3) to escape the system can the calculation be properly converged at all\cite{SRZ77}.  Here the density-driven error is more than 10 times larger than the magnitude of the functional error.  The density itself is very dramatically different from the exact one, and the PBE KS potential is not only too shallow, but is actually positive, and the eigenvalue is exactly zero.   This is a very abnormal calculation.  Evaluation of the PBE approximation on the exact density removes the density-driven error, and so drops the total energy error by an order of magnitude.  Even using the HF density is sufficient to
produce a much highly accurate electron affinity of H\cite{LFB10}.

We end this section by noting that the functional errors and density-driven errors have opposite signs. Thus, in a normal calculation, application of the approximate functional on the exact density will {\em increase} the energy error (albeit only slightly).  We have found this in all our calculations so far, which suggests this is typical behavior.  Thus we do {\em not} recommend universally using more accurate densities than the self-consistent density.  Only when a calculation is abnormal do we suggest such a procedure.


%


%

\subsection{History}

In the early days of DFT, it was often easier to use a HF code to find densities and evaluate XC approximations on those densities, again because such calculations were assumed to be normal\cite{B88, PCVJ92}. When DFT began to become popular in chemical applications in the early 1990's, this mode of testing approximations, called HF-DFT, was used in calculations\cite{GJPF92, OB94}. But very quickly, the computational and conceptual advantages of self-consistency led to self-consistent DFT calculations.  Moreover, the HF-DFT results were not systematically compared to self-consistent DFT calculations, except in some pioneering works which suggested that in difficult cases, the HF-DFT may yield more accurate answers\cite{S92}. The present paper may be considered as a fuller exploration and quantification of those early results.

\subsection{Context}

What does this analysis say about the many attempts to improve energetics in other ways, such as {\em ab initio} DFT\cite{VB12c}, random-phase approximation (RPA) \cite{F08, SHS08}, density-matrix functionals, etc? Our analysis explains several key features.  The first is that, despite having very wrong-looking XC potentials, and hence bad KS potentials, the effect of these errors on the energy via the density is minimal (a more detailed explanation is given in Ref. \onlinecite{BCL98}). Thus despite legions of papers reporting the very incorrect energy eigenvalues of KS potentials with approximate functionals, especially the highest occupied one which, for the exact functional, matches the negative of the ionization potential, DFT with these approximations continues to be very heavily used, because the energetics are unaffected.

Second, we now have a tool for quantifying the energetic error due to the density error. This allows us to ask questions such as when do we need to improve the density, in order to improve the energy. For example, DFT calculations using so-called exact exchange (EXX) have much more accurate KS potentials than those of standard approximations, yet usually worse energetics. This is because their functional errors usually outweigh the reduction in the density-driven error. Obviously, our present answer is a purely pragmatic one.  In specific circumstances, the density becomes sufficiently poor as to be the major source of error.  This shows that only in certain circumstances does this problem need to be addressed, and if ways could be found to avoid the poor self-consistent densities in such cases, a variety of apparent DFT errors would be avoided.

\section{HO$\cdot$Cl$^-$ complex}

\begin{figure*}[htb]
\begin{center}
\includegraphics[width=6.2in]{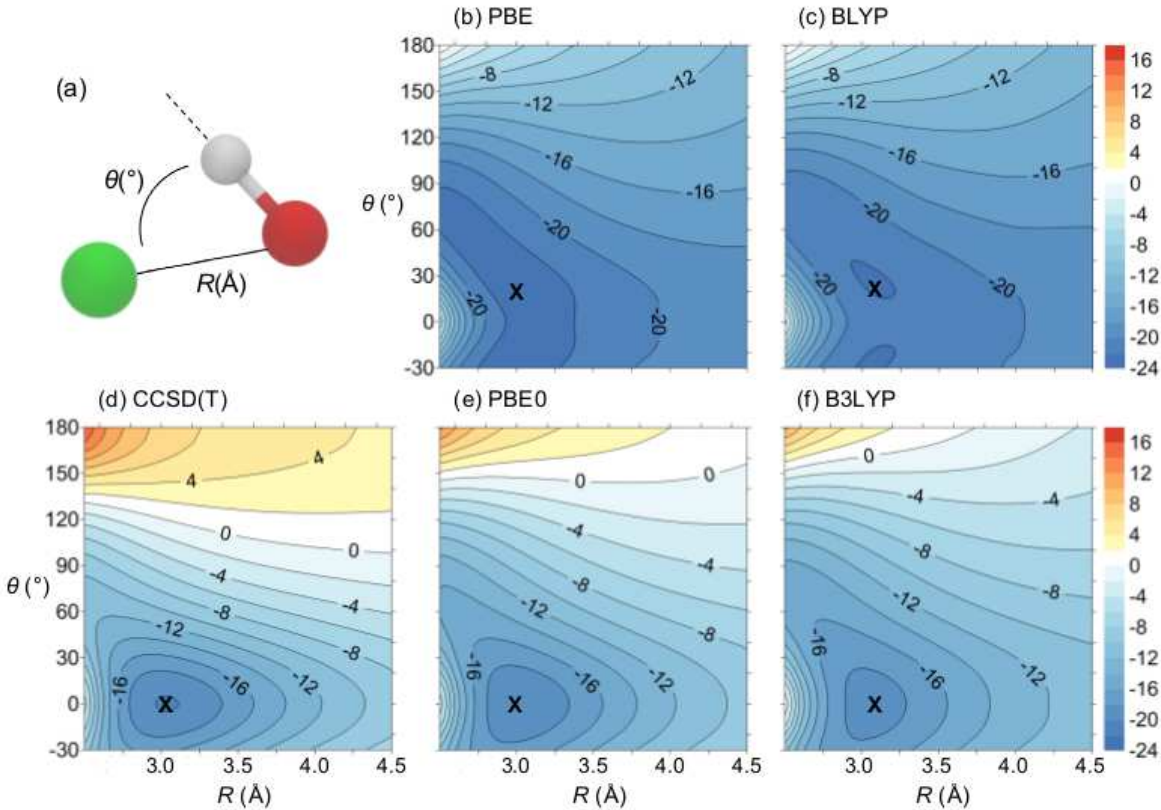}
\caption{PES of HO$\cdot$Cl$^-$ with various methods, using the AVTZ basis set. X marks the global minimum of each PES.}
\label{HOClPESfunc1}
\end{center}
\end{figure*}

We begin with the PES of the HO$\cdot$Cl$^-$ complex. The PES was scanned by changing the Cl-O distance ($R$) and Cl-O-H angle ($\theta$), as indicated in Fig. \ref{HOClPESfunc1}(a). The O-H bond length was fixed at 1.0 {\AA}. We performed single-point energy calculations with DFT, HF-DFT, and coupled-cluster method (CCSD(T)) on each geometry. We used GGA functionals (PBE\cite{PBE96} and BLYP\cite{B88, LYP88}), hybrid functionals (PBE0\cite{PEB96} and B3LYP\cite{B93}) and a double-hybrid functional with empirical dispersion correction (B2PLYP-D)\cite{G06, G06a}. We did these both in the gas phase and with implicit solvent. For implicit solvent calculations, we use the conductor-like screening model (COSMO)\cite{KS93} within the Turbomole suite\cite{turbomole}. We used Dunning's augmented correlation-consistent basis sets with X zeta functions (aug-cc-pVXZ, X = 2-3, denoted as AVXZ for simplicity from now on)\cite{D89, WD93}.

\subsection{Density functional approximation}

Contour plots of the HO$\cdot$Cl$^-$ complex PES evaluated with various methods are shown in Fig. \ref{HOClPESfunc1}. For this simple complex, we compare DFT results with CCSD(T), which we take as a benchmark. We notice several drastic failures of DFT approximation in these calculations. The worst qualitative failure is that the minimum of the GGA PES is not at 0$^\circ$, but is closer to 30$^\circ$. This is an incorrect hemi-bonding arrangement, attributed to the strong self-interaction error of the extra electron in the literature\cite{DKT08, VVS05, Y11}. We also note that the contours of the PES are quite incorrect in shape everywhere in the plane we have plotted.  Finally, the GGA PES is too negative overall (blue everywhere) indicating it is essentially useless for performing AIMD simulations of anions. While PBE is very popular for many materials simulations and static quantum chemical calculations, in fact, most AIMD simulations do not use PBE but other GGA's instead\cite{MTHP99}. This is because some key attributes of thermal simulations of water are incorrectly described by PBE. A popular alternative is BLYP, even though this is rarely used in regular quantum chemical calculations (unlike its hybrid off-spring, B3LYP). But a glance at Fig. \ref{HOClPESfunc1} (c) shows that BLYP is almost identical to PBE for this purpose, and suffers all the same difficulties.

On the other hand, hybrid density functionals, which add some fraction of HF to the GGA form, usually improve energetics\cite{KBP96} and partially correct self-interaction error.  So in Fig. \ref{HOClPESfunc1} (e) and (f) we plot the results with PBE0 and with B3LYP, which is the most popular functional in quantum chemistry.  We see that indeed there is great improvement. The minimum is now correctly at alignment, the surfaces are not entirely blue, and the shape is roughly correct.

As we have shown in Fig. \ref{HOClPESfunc2}, the results by evaluating DFT energies on HF densities are striking. In every case, we get essentially identical results throughout the plane of the PES.  All minima are in the correct locations, no curves are too blue, and all the details are correct.  The results are so consistent that we can draw several important conclusions.

\begin{itemize}

\item  Such good agreement confirms the theory behind DC-DFT.  All these DFT calculations are {\em abnormal}, and the error is greatly reduced by using a better density.  It also confirms that the HF density is sufficiently more accurate than the self-consistent DFT density for these calculations to produce much more accurate energies.

\item For this problem, we no longer need the benchmark defined by CCSD(T). The extreme level of consistency between so many different DFT approximations imply that all such calculations are yielding a very accurate answer.

\item While the hybrid functionals definitely improve over the GGA's, the primary effect is the improvement in the self-consistent density due to the HF component in the energy.  In fact, evaluated on a sufficiently accurate density, there is no need to use a hybrid functional (but of course, finding the HF density is relatively expensive in AIMD calculations).

\end{itemize}

\begin{figure}[htb]
\begin{center}
\includegraphics[width=2.5in]{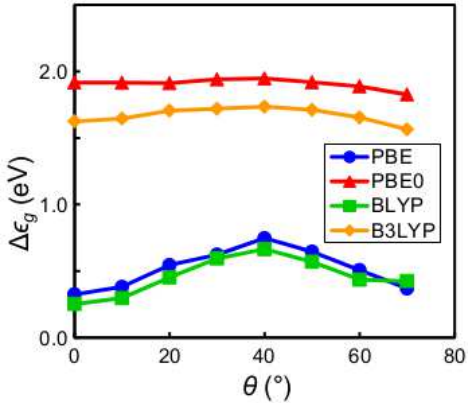}
\caption{KS HOMO-LUMO gap, $\deps$, of HO$\cdot$Cl$^-$ complex for several approximate functionals.} 
\label{HOClgapfunc}
\end{center}
\end{figure}

\begin{figure*}[htb]
\begin{center}
\includegraphics[width=6.5in]{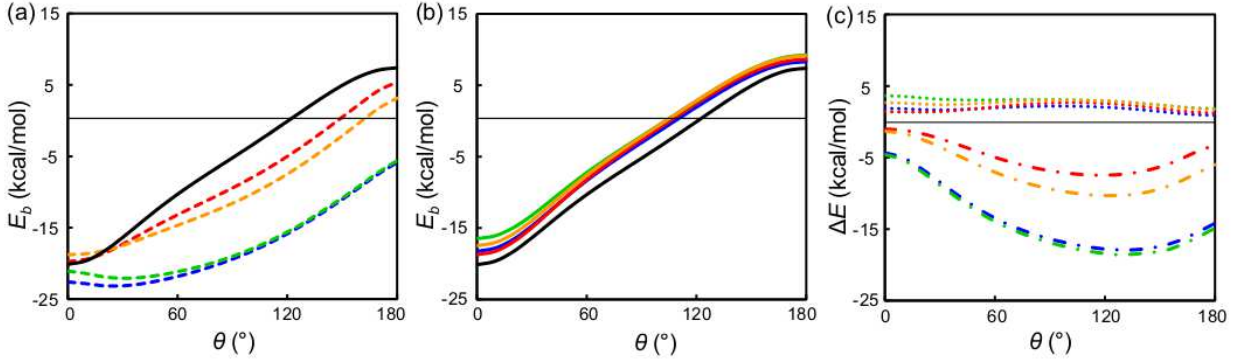}
\caption{Binding energy along $R$ = 3.0 {\AA} of HO$\cdot$Cl$^-$: CCSD(T) (black), PBE (blue), BLYP (green), PBE0 (red), B3LYP (orange). Panel (a) shows self-consistent results, (b) are HF-DFT results, while (c) shows $\Delta E_D$ (dot-dash) and $\Delta E_F$ (dotted). AVTZ basis set is used in all calculations.}
\label{HOClfunc30R}
\end{center}
\end{figure*}

Next we consider the gap in the approximate self-consistent KS calculations. Part of the DC-DFT theory is that an abnormal system should have an unusually small KS gap, suggesting that its density is unusually inaccurate\cite{KSB13}. Fig. \ref{HOClgapfunc} shows $\deps$ for the HO$\cdot$Cl$^-$ complex. Each point corresponds to the energy minimum of each $\theta$, i.e., using $R$ that gives the lowest binding energy for given $\theta$. Since we scanned $R$ = 2.5 $\sim$ 4.5 {\AA} for calculations, we excluded $\theta >$ 70$^\circ$ from this figure where the energy minimum was located at $R$ = 2.5 {\AA} or $R$ = 4.5{\AA}. The GGA methods clearly have small $\deps$ (less than 1 eV) which is consistent with the large density-driven error in these methods. In the case of hybrid methods, the $\deps$ is a mixture of a HF gap and a KS gap rather than pure KS gap, so it may not be as good as an indicator for density-driven errors. The hybrid $\deps$ are not as small as in GGA, but still less than 2 eV, which explains the moderate density-driven error compared to GGA methods. As some DFT calculations with $\deps$ even as high as 2.5 eV have large density-driven error\cite{KSB13}, calculations with $\deps$ below 2 eV should be suspected of being abnormal. 

To gain more insight and quantitative understanding, we show energy curves along $R$ = 3.0 {\AA} in Fig. \ref{HOClfunc30R}. The binding energy $E_b$ is defined as $E_b = E[\mbox{HO}\cdot\mbox{Cl}^{-}] - (E[\cdot\mbox{OH}] + E[\mbox{Cl}^{-}]$), where $E[\mbox{HO}\cdot\mbox{Cl}^{-}]$, $E$[$\cdot$OH], and $E$[Cl$^{-}$] is the energy of HO$\cdot$Cl$^{-}$ complex, OH radical, and Cl$^{-}$ anion respectively. We see very clear patterns.  In Fig. \ref{HOClfunc30R}(a), the GGA's (blue and green) produce incorrect minima at $\theta$ = 20$^\circ$. Hybrid methods (red and orange) are much more accurate near the minimum, but show increasing error as $\theta$ gets larger. On the other hand, in Fig. \ref{HOClfunc30R}(b), all the GGA and hybrid curves line up almost perfectly, once evaluated on HF densities.  The small remaining deviation among them is near the minimum, where the PBE methods (PBE and PBE0) are most accurate.  In any case, all are slightly shifted above the accurate CCSD(T) curve.


Finally, we decompose the energy error into density-driven and functional errors for both the GGAs and hybrids in Fig. \ref{HOClfunc30R}(c). GGA methods show large density-driven error for all regions, maximizing at $\theta$ = 130 $\sim$ 140$^\circ$, while hybrid methods have less but still significant density-driven error maximizing at $\theta$ = 120 $\sim$ 130$^\circ$. Functional errors stay almost constant for every $\theta$. The evaluated density changes as the geometry changes and the functional used is left unchanged, resulting in this independence of functional error to geometry. 


\subsection{Basis sets}

%

\begin{figure*}[htb]
\begin{center}
\includegraphics[width=5.0 in]{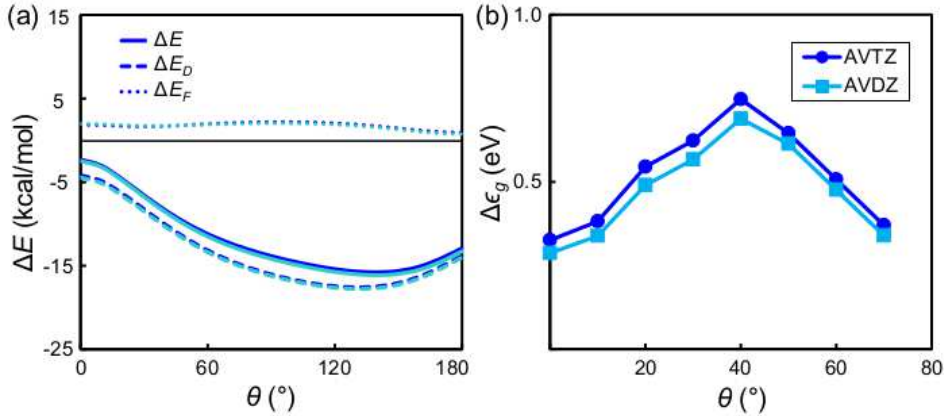}
\caption{(a) Decomposition of HO$\cdot$Cl$^-$ complex energy curve along $R$ = 3.0 {\AA} of PBE/AVDZ (light blue) and PBE/AVTZ (blue) calculations. (b) $\deps$ of HO$\cdot$Cl$^-$ complex for each basis set.}
\label{HOClbas30ang2}
\end{center}
\end{figure*}

We have also calculated PES for both PBE and B3LYP, self consistently and in HF-DFT, using a smaller basis, namely, AVDZ.  These are qualitatively and quantitatively almost identical to those with AVTZ, showing basis set convergence, and that AVDZ may be sufficient for most purposes for these calculations.  This is illustrated in the energy error decomposition for PBE in Fig. \ref{HOClbas30ang2}(a), where the shifts from one basis to the next are tiny compared to all other energy error contributions. The $\deps$ of both methods in Fig. \ref{HOClbas30ang2}(b) are also similar.

\subsection{B2PLYP-D functional}

\begin{figure*}[htb]
\begin{center}
\includegraphics[width=6.2in]{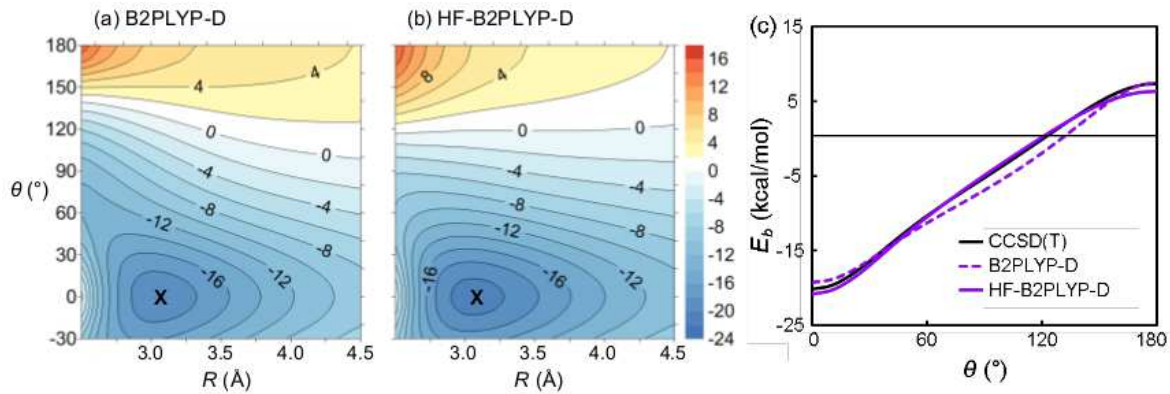}
\caption{PES of HO$\cdot$Cl$^-$ with the (a) B2PLYP-D functional and (b) HF-B2PLYP-D method, with its (c) binding energy curve along $R$ = 3.0 \AA. The AVTZ basis set is used in all calculations. X marks the global minimum of each PES.}
\label{HOClPESb2plyp}
\end{center}
\end{figure*}

Next, we show what happens when we use a more modern and more accurate approximate functional for this problem. The B2PLYP-D functional is a double hybrid functional combined with empirical dispersion parameters\cite{G06, G06a}. In conventional hybrid functionals, HF exchange is added as the non-local exchange contribution.  In addition to this, B2PLYP-D has the non-local perturbation correction added for the correlation part by second-order perturbation theory. This is based on {\em ab initio} Kohn-Sham perturbation theory (KS-PT2) by G\"{o}rling and Levy\cite{GL93, GL94}. Due to the large Fock exchange fraction, self-interaction error is greatly reduced, while the side effects of having large Fock exchange, such as incomplete static correlation, are alleviated by the second-order perturbation in the correlation\cite{SG08}. This leads to excellent results in many cases\cite{SG07, SG08}, including two-center three-electron bonding in radical complexes\cite{DLAJ11}, which makes the method a great choice of benchmark for this work.

 In Fig. \ref{HOClPESb2plyp}(a), this approximation is doing an excellent job of reproducing the PES everywhere, on its {\em self-consistent density!} Thus this functional is sufficiently accurate for this problem that this is a {\em normal} calculation. Sure enough, when we repeat the calculation using the HF density, as shown in Fig. \ref{HOClPESb2plyp}(b), the PES worsens. This strongly suggests that the B2PLYP-D self-consistent density is better than the HF density here. Thus this functional can be used for this problem without modification, so long as the user can afford to evaluate it, and should not be density corrected.  But all the cruder older approximations yield abnormal results and need correction.







\subsection{Solvation}

\begin{figure*}[htb]
\begin{center}
\includegraphics[width=4.3 in]{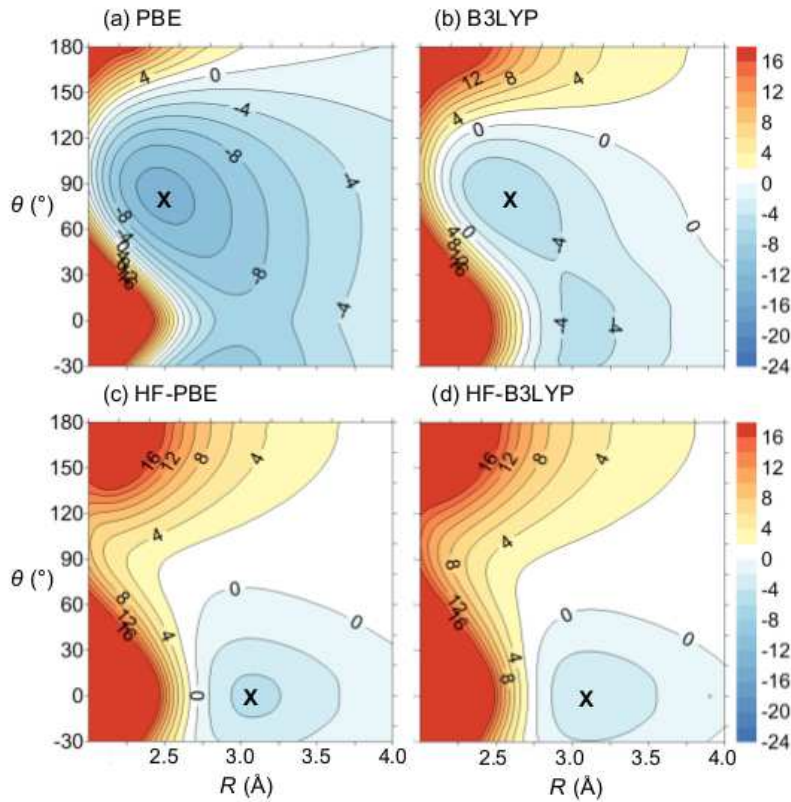}
\caption{PES of HO$\cdot$Cl$^-$ with implicit water in contour plot. PES is constructed upon (a) PBE, (b) B3LYP, (c) HF-PBE, and (d) HF-B3LYP scanning results. AVTZ basis set used in all calculations.}
\label{HOClPESsolv}
\end{center}
\end{figure*}

D'Auria {\em et al.}\cite{DKT08} performed AIMD simulations on the HO$\cdot$Cl$^-$ complex in explicit water solvents where the minimum appeared to be a hemi-bonding structure ($\theta$ = 80$^\circ$) with a standard approximate functional BLYP, while self-interaction corrected BLYP (BLYP-SIC) gave a hydrogen-bonding minimum structure on $\theta$ = 0$^\circ$. They then scanned the gas phase HO$\cdot$Cl$^-$ complex PES along $\theta$ = 0$^\circ$ and $\theta$ = 80$^\circ$ based on those observations. As we observed in our gas phase calculation, the true minimum of gas phase HO$\cdot$Cl$^-$ complex lies somewhere between $\theta$ = 0$^\circ$ and 80$^\circ$. To look in more depth at solvation effects, we show contour plots of HO$\cdot$Cl$^-$ complex in implicit water solvent in Fig. \ref{HOClPESsolv}.

 For both PBE and B3LYP calculations, the minimum is clearly a hemi-bonding structure with $\theta$ = 80$^\circ$, in contrast to the gas phase calculation, where the global minimum was at $\theta$ = 20$^\circ$. The Fock exchange in the hybrid functional indeed has some effect, producing a second local minimum along the hydrogen-bonding region ($\theta$ = 0$^\circ$), yet did not correct the overstabilization of the hemi-bonding structure, resulting in the wrong global minimum.

 On the other hand, the sole minimum of both DC-DFT calculations is the hydrogen-bonding structure for both gas phase and implicit water calculations. Unlike the self-consistent DFT results, the PES are quite similar regardless of functional, which was a trait also observed in gas phase PES, and no sort of local minimum is shown in the hemi-bonding region. This shows the accuracy of DC-DFT is on par with SIC-DFT with far less computational cost, at least for non-periodic cases, even in the presence of implicit solvent. Finally, we would like to mention that there are some works that show KS-DFT greatly underestimates the redox potential of OH$\cdot$/OH$^-$ and Cl$\cdot$/Cl$^-$ in explicit solvent while this is not the case in implicit solvent simulations\cite{ACCS12}. Even though it is possible that there is some difference between our results and the explicit solvent results, our findings matched the explicit solvent result unlike the OH$\cdot$/OH$^-$ and Cl$\cdot$/Cl$^-$ case. We believe this discrepancy is due to the difference between a anion-radical complex and a lone radical, alleviating the effect of extended states of explicit solvent.























\section{HO$\cdot$H$_2$O complex}

\begin{figure}[htb]
\begin{center}
\includegraphics[width=4.3in]{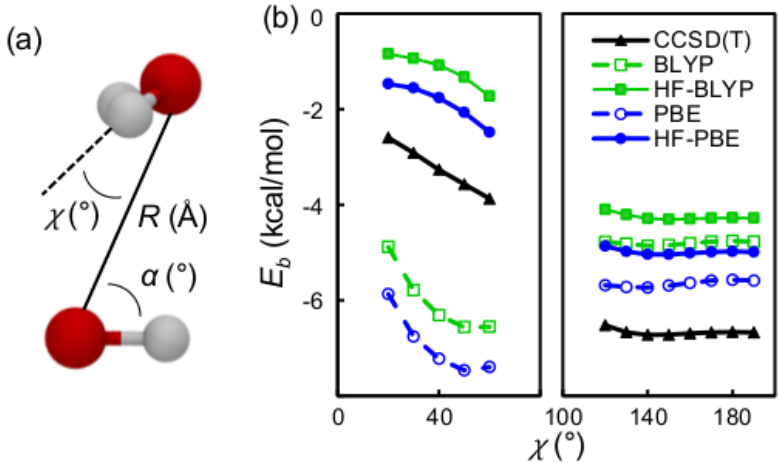}
\caption{(a) Geometrical parameters used in calculations of HO$\cdot$H$_{2}$O complex. (b) Comparison of the PES scan for HO$\cdot$H$_2$O complex using various methods. Binding energies are plotted against $\chi$. Each point is using the minimum energy geometry for given $\chi$.}
\label{HOH2OPEScurve}
\end{center}
\end{figure}

\begin{figure*}[htb]
\begin{center}
\includegraphics[width=4.1in]{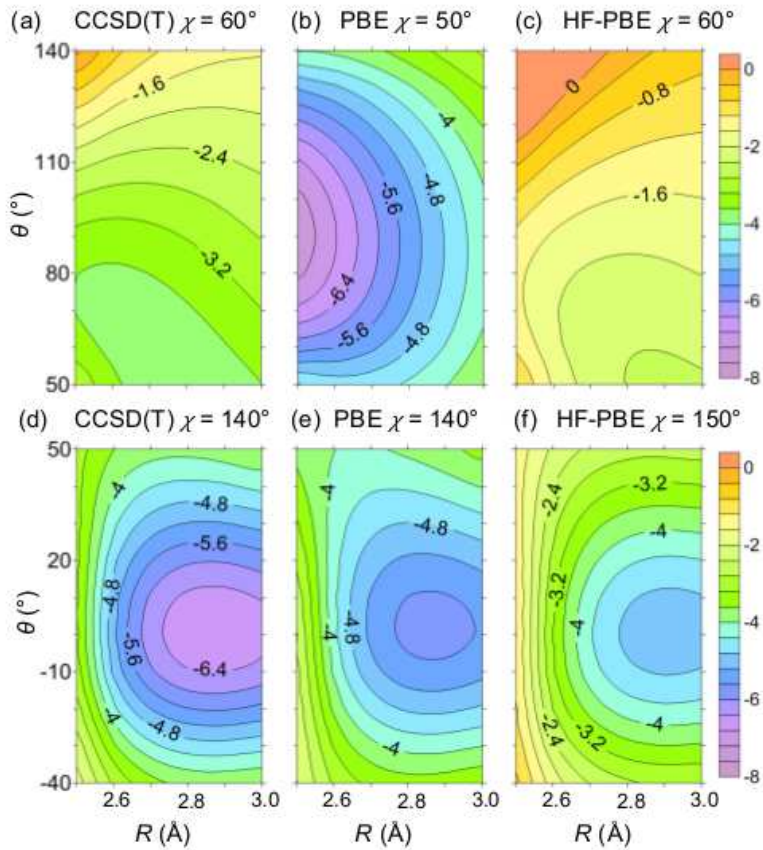}
\caption{PES of HO$\cdot$H$_2$O with various methods for specific $\chi$ values. The upper panels correspond to hemi-bonding structures, while the lower panels correspond to hydrogen-bonding structures. AVTZ basis set used for all calculations.}
\label{HOH2OPES}
\end{center}
\end{figure*}

\begin{figure}[htb]
\begin{center}
\includegraphics[width=3in]{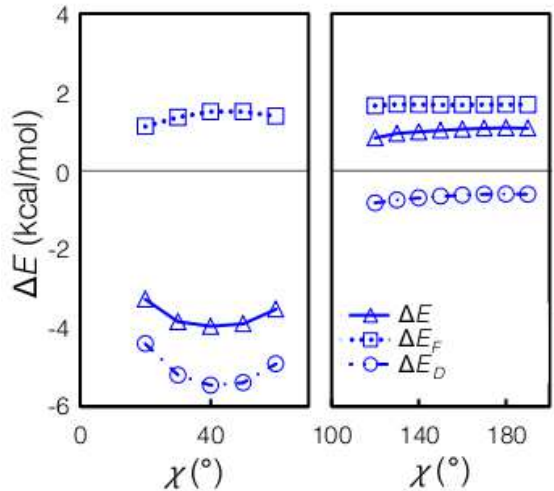}
\caption{Error decomposition for PBE PES scans of HO$\cdot$H$_2$O complex. Each point is using the minimum energy geometry for given $\chi$.}
\label{HOH2Oerror}
\end{center}
\end{figure}

 To confirm that the performance of DC-DFT is not restricted to anion complexes, we also look at a neutral radical complex. We evaluated the PES of the HO$\cdot$H$_2$O complex using DFT, HF-DFT and CCSD(T) with AVTZ basis set. PBE and BLYP functionals are used in DFT calculations. We used the same parameters used by Chipman\cite{C11}, depicted in Fig. \ref{HOH2OPEScurve}(a). The evaluated PES is depicted in Figs. \ref{HOH2OPEScurve} and \ref{HOH2OPES}. The binding energy $E_b$ here is defined as $E_{b} = E[$HO$\cdot$H$_2$O] - ( E[$\cdot$OH] + E[H$_2$O]), where $E[$HO$\cdot$H$_2$O], $E[\cdot$OH], and $E[$H$_2$O] is the energy of the HO$\cdot$H$_2$O complex, OH radical, and H$_2$O molecule, respectively.

 In Fig. \ref{HOH2OPEScurve}, each point indicates the minimum energy possible for a given $\chi$. We chose to scan between $\chi = 20^{\circ}$ to $60^{\circ}$ and $120^{\circ}$ to $190^{\circ}$. The $\chi = 20^{\circ}$ to $60^{\circ}$ region is where the hemi-bonding geometry was observed in excited state calculations, while the latter is where the global minimum of the ground state was discovered\cite{C11}. 

 The CCSD(T) results reproduce Chipman's result, where the global minimum is in the hydrogen-bonding region of $\chi = 140^{\circ}$. On the other hand, despite having a minimum in the hydrogen region, both DFT methods clearly have the global minimum in the hemi-bonding region of $\chi = 50^{\circ}$. Now, as one can expect from the HO$\cdot$Cl$^-$ complex results, the HF-DFT curve successfully resembles the CCSD(T) results, having the global minimum in the hydrogen-bonding region of $\chi = 150^{\circ}$. Also CCSD(T) and HF-DFT results have no minimum in the hemi-bonding region, so the $\chi$ value with the lowest energy is 60$^{\circ}$.

 We scanned through $R$ and $\alpha$ on the $\chi$ values that give minimum energy for each method and region in Fig. \ref{HOH2OPES}. Once again, the PES of HF-PBE looks like the PES of CCSD(T) with an energy shift in both hemi-bonding and hydrogen-bonding region, while PBE has a clearly different PES in the hemi-bonding region.

Fig. \ref{HOH2Oerror} shows the error decomposition of the PBE calculation.  As expected, calculations in the hemi-bonding region exert a strong density-driven error. In the hydrogen-bonding region, the density-driven error is quite small compared to the hemi-bonding region, but $\deps$ in both regions is still quite small ($\deps$ = 1.09 eV at the hemi-bonding minimum, 0.97 eV at the hydrogen-bonding minimum).




\section{Conclusion}

Approximate DFT typically suffers from severe self-interaction error in calculations of odd-electron radical complexes\cite{DKT08, VVS05}. This work shows that the density correction in DC-DFT, even using simple HF densities, can often give more accurate results than DFT using self-consistent densities in these types of calculations. To explain this in a systematic way, we showed {\em any} approximate calculations can be classified into one of the two types. In {\em normal} calculations, the functional error dominates, while in {\em abnormal} calculations, the density-driven error is larger than the functional error. We illustrated this using simple two-electron systems, namely H$^-$ anion and He atom. Approximate DFT was abnormal in H$^-$ showing severe density-driven errors, while all other cases were normal. In these normal cases, the density-driven error was negligible, despite having very wrong-looking KS potentials. By this analysis, we stressed that DC-DFT is likely to give better results than approximate DFT with self-consistent densities only for abnormal calculations, and not for normal calculations. We presented PES' of an anion radical complex, e.g., the HO$\cdot$Cl$^-$ complex, using various common approximate functionals including GGA functionals and hybrid functionals and even more modern functionals like B2PLYP-D, with both self-consistent DFT and DC-DFT. Both GGA and hybrid functionals behaved poorly self consistently and only using the highly accurate B2PLYP-D self-consistent density was sufficient for getting accurate PES. On the other hand, DC-DFT gave identical PES' regardless of the approximate functional used, and gave correct global minima and PES slopes, showing GGA and hybrid approximate functional calculations are abnormal calculations. Using the concept introduced from our previous letter\cite{KSB13}, we showed a very small $\deps$ can be used as an indicator of abnormality. We checked the calculations with different basis sets where we have similar results for both AVDZ and AVTZ basis set. We also showed even calculations with implicit solvent have similar tendencies, where self-consistent GGA and hybrid densities give poor results, predicting hemi-bonding structures as the global minimum. PES' evaluated from DC-DFT were once more identical, independent of the functional used. Finally, we examined the validity of DC-DFT for neutral radical complexes by evaluating PES' of HO$\cdot$H$_2$O. Self-consistent DFT predicted the hemi-bonding structure as the global minimum, while DC-DFT correctly predicted the hydrogen-bonding structure as the global minimum. These results and the $\deps$ showed the abnormality in these self-consistent DFT calculations. DC-DFT can be used as a simple cure of abnormality, i.e., strong density-driven error especially driven from self-interaction, which has less computational cost and is free of empirical parameters compared to various SIC methods. We must mention the HF density we used in this work may not be appropriate for all cases, including cases with strong spin-contamination\cite{KSB11}, or periodic boundary conditions. Nonetheless, one can use DC-DFT with any other source of accurate densities in cases where the HF density is not suited. Additionally, we expect this method to give promising results for various problems that are challenging for approximate DFT like reaction barriers and dissociation\cite{KSB13}, when the errors are density-driven.


\setcounter{secnumdepth}{0}

\section{Acknowledgements}

We thank Eberhard Engel for the use of his atomic OEP code and Cyrus Umrigar for the exact two electron system data from QMC calculation. This work was supported by the global research network grant (No. NRF-2010-220-C00017) and the national research foundation [2012R1A1A2004782 (E. S.)]. K.B. acknowledges support under NSF CHE-1112442. M-C thanks the fellowship of the BK 21 program, and Sanghyeon Lee for his effort.

\vspace{0.5cm}\rule{\linewidth}{0.5mm}\vspace{0.2cm}

* Corresponding Author: esim@yonsei.ac.kr \vspace{0.5mm}


\sf 
\label{page:end}



\clearpage

%

\end{document}